\documentclass[10pt,twocolumn,a4paper]{article}

\usepackage{times}
\usepackage[font=small,labelfont=bf]{caption}
\usepackage[top=20mm,left=20mm,right=20mm,bottom=25mm]{geometry}

\usepackage{cite}
\usepackage[utf8]{inputenc}
\usepackage[T1]{fontenc}
\usepackage{amsmath,amssymb,amsfonts}
\usepackage{algorithmic}
\usepackage{graphicx}
\usepackage{textcomp}
\usepackage{xcolor}
\usepackage{hyperref}
\usepackage{comment}

\usepackage{enumitem}

\begin{document}

\title{\Huge{Ethical Interviews in Software Engineering}}
\date{\small{Accepted to ESEM'19: International Symposium on Empirical Software Engineering and Measurement 2019}}

\author{Per Erik Strandberg\\
  \small{Westermo Network Technologies AB, Västerås, Sweden, per.strandberg@westermo.se} \\
  \small{Mälardalen University, Västerås, Sweden, per.erik.strandberg@mdh.se}
}

\maketitle

\newcommand{\mychecklist}[1]{\subsubsection*{Check\-list for #1}}
\newcommand{\myexample}[1]{\begin{quote}\emph{#1}\end{quote}}

\newenvironment{IEEEkeywords}
{\vspace{3mm}\noindent \emph{Index Terms} -- }

\renewenvironment{abstract}
{\section*{Abstract}}

\renewcommand{\thesubsection}{\Alph{subsection}}

\begin{abstract}
\noindent
\emph{Background:}
Despite a long
history, numerous laws and regulations, ethics remains an unnatural
topic for many software engineering researchers.
Poor research ethics may lead to mistrust of research
results, lost funding and retraction of publications.
A core principle for research ethics is confidentiality,
and anonymization is a standard approach to guarantee it.
Many guidelines for
qualitative software engineering research, and for qualitative research
in general, exist, but these do not penetrate how and why to
anonymize interview data.
\\ \noindent \emph{Aims:}
In this paper we aim to identify ethical guidelines
for software engineering interview studies involving industrial practitioners.
\\ \noindent \emph{Method:}
By learning from previous experiences and listening to the
authority of existing guidelines in the more mature field of
medicine as well as in software engineering,
a comprehensive set of checklists for interview studies was distilled.
\\ \noindent \emph{Results:}
The elements of an interview study
were identified and ethical
considerations and recommendations for each step were produced, in
particular with respect to anonymization.
Important ethical principles are:
consent, beneficence, confidentiality, scientific value, researcher
skill, justice, respect for law, and ethical reviews.
\\ \noindent \emph{Conclusions:}
The most important contribution of this study is the
set of checklists for ethical interview studies.
Future
work is needed to refine these guidelines with respect to legal
aspects and ethical boards.
\end{abstract}


\section{Introduction}


When humans are data sources for researchers,
the research community has to respect them and protect them from harm.
However, this has not always been, and still is not always, the case
in research.
A study published in early 2019 investigated questionable ethics in
research on Chinese transplant recipients \cite{rogers2019,caplan2011}.
The authors request the
immediate retraction of a large body of papers
because of the poor ethical principles behind the studies
(e.g., using organs from executed prisoners).
In addition to physical risks like disease or death, researchers might
cause other types of harm, such as risks with respect to
privacy, personal values, or family links
(e.g., if they expose illegal, sexual or deviant behavior)
\cite{carusi2009}.
Related to the field of software engineering, many
data sources may cause harm to an individual:
exposure of financial data, message history or dating app logs
are some examples.
When industry practitioners participate in research, they might
expose information that could cause harm not
only to individuals, but also companies,
e.g.,\ by mentioning quality shortcomings.
Software has a profound impact
on almost every aspect of society, making ethics in software
engineering research an important topic.

A common research method for data collection 
is the \emph{interview}.
Singer et al.\ describe the interview as a method
where at least one researcher talks to at least one interviewee \cite{singer2008}.
Two common approaches to interviews are structured and semi-structured
interviews: in the former the researcher asks all questions in the same way,
and in the same order, to all interviewees.
In the latter some flexibility is allowed.
Interviews may be time consuming, but are suitable
for many types of research methods and philosophical traditions.
%
%

Many guidelines exist for empirical software engineering research,
such as the books by
Kitchenham et al.\ \cite{kitchenham2015},
Runeson et al.\ \cite{runeson2012}, and
Shull et al.\ \cite{shull2008}.
However, 
these often lack
how to conduct interviews and handle
interview artifacts with respect to ethical considerations.
In this paper we
fill this gap by
continuing the tradition to transfer guidelines
from medicine into the field of software engineering.
First, in Section~\ref{background-and-related-work} 
we review existing guidelines for ethical research,
guidelines for software engineering,
and revisit an interview study we recently finalized. 
Second, in Section~\ref{interview-life-cycle}
we investigate the research interview
and consider ethical aspects in each element.
This section contains the main contribution of this paper:
guidelines in the form of check\-lists,
that simplifies ethical research, in particular with respect to
how to practically anonymize and work with interview data.
Finally, in Sections~\ref{summary} and~\ref{conclusion} we
summarize, discuss and conclude this paper.

\begin{table*}[th!]
\begin{center}
\begin{tabular}{lp{5.5in}}
Ethical Principle& Summary \\
\hline
Consent          &
Participation should be voluntary, and
withdrawal possible at any time.
Participants should be informed of this in a way
that they can understand.
\\
Beneficence      &
The welfare of participants, and the greater good for society,
should be considered.
\\
Confidentiality  &
The privacy and confidentiality of the participants
must be protected in order to minimize the impact of the study on
their integrity. 
\\
Scientific Value &
Research should yield fruitful results for the
good of society, and not be random and unnecessary. 
\\
Researcher Skill &
The researchers should have adequate skills.
\\
Justice          &
It is injust to let one group carry the burden of research
while another gets the benefits of research. 
\\
Respect Law      &
Relevant laws should be obeyed. 
\\
Ethics Review    &
An independent ethics board should comment on, guide and approve
studies involving humans.
\\
\end{tabular}
\caption{Ethical principles in research. \label{tab-eth-princ}
}
\end{center}
\end{table*}

\section{Background and Related Work}
\label{background-and-related-work}

This section covers a chronology of
ethical principles in medicine,
guidelines in software engineering,
anonymization, as well as
legislation and institutional review boards.
Important ethical principles highlighted in previous and related works
are:
consent,
beneficence,
confidentiality,
scientific value,
researcher skill,
justice,
respect law, and
ethics review
(summarized in Table~\ref{tab-eth-princ}).
Despite much previous work, how to apply
these principles for interviews in software engineering research
has remained unexplored.

\subsection{Guidelines from Cos to Menlo via Nuremberg and Helsinki}
\label{background}


Medical research has a long history of
guidelines. Some 2500 
years ago Hippocrates of Cos wrote the Hippocratic oath \cite{hippo1923}.
A core topic is beneficence, 
often misquoted as ``first, do no harm'' \cite{smith2005}.

Following the second world war and the monstrous Nazi experiments on humans,
ten ethical research principles constitute
the \emph{Nuremberg Code} \cite{nuremberg1947}.
The principles highlight that
an experiment should aim for positive results for society,
participants must consent to and have the right to withdraw from a study,
risks should be minimized and not exceed the expected benefits,
the experiment must stop if continuation would be dangerous,
research should be based on previous knowledge, and that
staff must be qualified to conduct the experiment.
In the similar declarations of Helsinki \cite{helsinki2014}
and later Taipei \cite{taipei2016},
the World Medical Association (WMA) laid out
ethical considerations regarding medical research involving human subjects
as well as health databases and biobanks.

In practice, a way to respond to
unethical research is to create ethical guidelines.
One example is the Belmont report \cite{belmont1978} that
strives to protect human subjects.
This report points out 
principles for research involving humans 
and the importance of a risk/benefit assessment,
where stakeholder identification is prerequisite.
It argues that the justice principle has
implications for the selection of participants in a study:
it is not fair if one group takes the burden of
research whereas another receives the benefits.
Another response to unethical research is for
society to respond with laws governing research,
and with new laws there is an increased need for more
guidelines. 
Laws on public records, corporate secrets, public archiving,
etc., might, from the researcher perspective,
seem to be in conflict with laws covering ethical research.


In the Menlo report \cite{menlo2012},
the Belmont report is adapted and built upon
for the field of Information and Communication Technology (ICT)
and ICT Research (ICTR).
It reiterates the core principles from the Belmont report,
and adds the additional principle of
respect for law and public interest.
The Menlo companion \cite{menlo2013}, aims at providing more
concrete guidelines for ethical ICTR.
Together they argue that ICTR is different from medical research
because of the greater distances between researcher and participant,
as well as the scale, speed, wide distribution, and opacity of the field.
In ICT there is also the potential for collecting many types of data --
from a cell phone one might collect
financial and geographical data, as well as emails, dating history, etc.

\label{related-work}

\subsection{Guidelines in Software Engineering}

Three examples of books with guidelines for
empirical software engineering research
are
Kitchenham et al.\ \cite{kitchenham2015},
Runeson et al.\ \cite{runeson2012}, and
Shull et al.\ \cite{shull2008}.
They all highlight the importance of ethics,
however they lack
hands-on instructions on how to conduct interviews and
how to handle interview artifacts with respect to ethical perspectives.
Vinson and Singer (in one of the chapters in Shull et al.)
repeat four ethical principles from medicine:
informed consent,
beneficence,
confidentiality,
and scientific value
\cite{vinson2008}.
Kitchenham et al.\ highlight ethical issues
for primary studies,
in particular with respect to informed participation,
pressure to take part, collecting demographic data
and reporting \cite{kitchenham2015}.

The recent ACM Code of Ethics 
\cite{acm2018}
reiterates a number of the already mentioned guidelines.
Despite the enormous scope of the code, a number of the
principles could be seen as relevant for
interview research in software engineering, in particular
\emph{respect privacy},
and \emph{honor confidentiality}
that, again, emphasize the importance of protecting the interviewees.
\emph{Accept and provide appropriate professional review},
could be seen as encouragement for a review step where the interviewee
may comment on the transcript.
We are also reminded of the Nuremberg code in:
\emph{Perform work only in areas of competence}.

\subsection{Anonymization}

Researchers have both ethical and legal obligations to respect the
confidentiality of individuals,
this confidentiality
is also essential for maintaining trust
\cite{taipei2016}.
A good way to mitigate ethical concerns
is to anonymize personal data.
If a link is needed between participants and
personal data, then details for the link could be
pseudonymized \cite{eudata2018}.

Becker-Kornstaedt describes how interview data
could be handled during software process modeling
\cite{becker2001}.
She describes ethical dilemmas and techniques
to protect the interviewees.
Mitigations include
anonymization of the data
and
sanitation of data, where certain details are
left out, summarized or aggregated. However, sanitized
data might be too abstract or unsuitable for the task.
She also proposes to
be transparent about risks,
have a well defined scope such that data collection can be limited
and irrelevant data left out.
She suggests to
make sure that managers are not present during interviews,
to avoid giving away raw data such as interview transcripts,
and
to allow the interviewee to review data with respect to completeness as
well as confidentiality.

Aldridge et al.\ discuss the problem of data proliferation.
They conduct a type of life cycle analysis of an interview,
and propose 14 guidelines for data security \cite{aldridge2010}.
They recommend early anonymization, because a leak of anonymized data
reduces harm when compared to loss of not yet anonymized data.
They recommend replacing identifiers such as names, places or organizations
with unique identifiers (pseudonyms).

Saunders et al.\ \cite{saunders2015b, saunders2015a}
cover ethical interviews in the field of medicine,
and also propose pseudonymization, e.g.,\ by replacing
the role of a mother to that of a sister
during the transcription.

In our study we
interviewed industrial practitioners
on the flow of information in software testing \cite{strandberg2019}.
We anonymized the transcripts.
The most anonymized categories were
names of tools or tool jargon,
names or details of products and organizations, and
extracts related to the domain.

\subsection{De-anonymization}

De-anonymization, or data re-identification, refers to the practice of
uncovering the identify of an individual by using anonymized data.
For example, Sweeney 
showed that 87\% of all Americans could be uniquely identified by
their ZIP code, gender, and date of birth \cite{sweeney2000}.
Rosenblum et al.\ %
conducted a study where they identified source code authors
by investigating compiled binaries \cite{rosenblum2011}.

Saunders et al.\ 
describe
their experiences from research where they interviewed family members
of patients with severe brain injury, vegetative or minimally conscious states,
and provide guidelines for ethical interviews
\cite{saunders2015b, saunders2015a}.
Their interviewees are from a small sample,
some participants have been part of court cases or covered in the media,
and some are active in social networks
such as blogs or support forums.
If an interview covers a topic already mentioned in a
blog or court case, then de-anonymization may be trivial for a
motivated reader
if details are also retold in a publication.

\subsection{Drawbacks of Anonymization}

Nespor investigates anonymization
of locations in qualitative research \cite{nespor2000}.
He argues that desire to anonymize comes from three assumptions:
(i) Identification may cause harm -- an assumption he finds plausible.
(ii) Anonymization decreases the likelihood of identification -- an
assumption he claims lacks support in research.
(iii) Identifying places and settings make participants more easily
identifiable -- an assumption he says might be relevant.
When reading Nespor it is clear that anonymization is related to
generalizability: results from a certain high school, in a certain
place at a certain time might be part of a complex interplay with
the local community, the history of the school, etc. 
Results from ``a high school'' might instead imply that
the results apply in general for all high schools.

Given the importance of anonymization in guidelines and laws, we will
not further investigate drawbacks of anonymization in this paper.

\begin{figure*}[th!]
  \begin{center}
  \includegraphics[width=\textwidth]{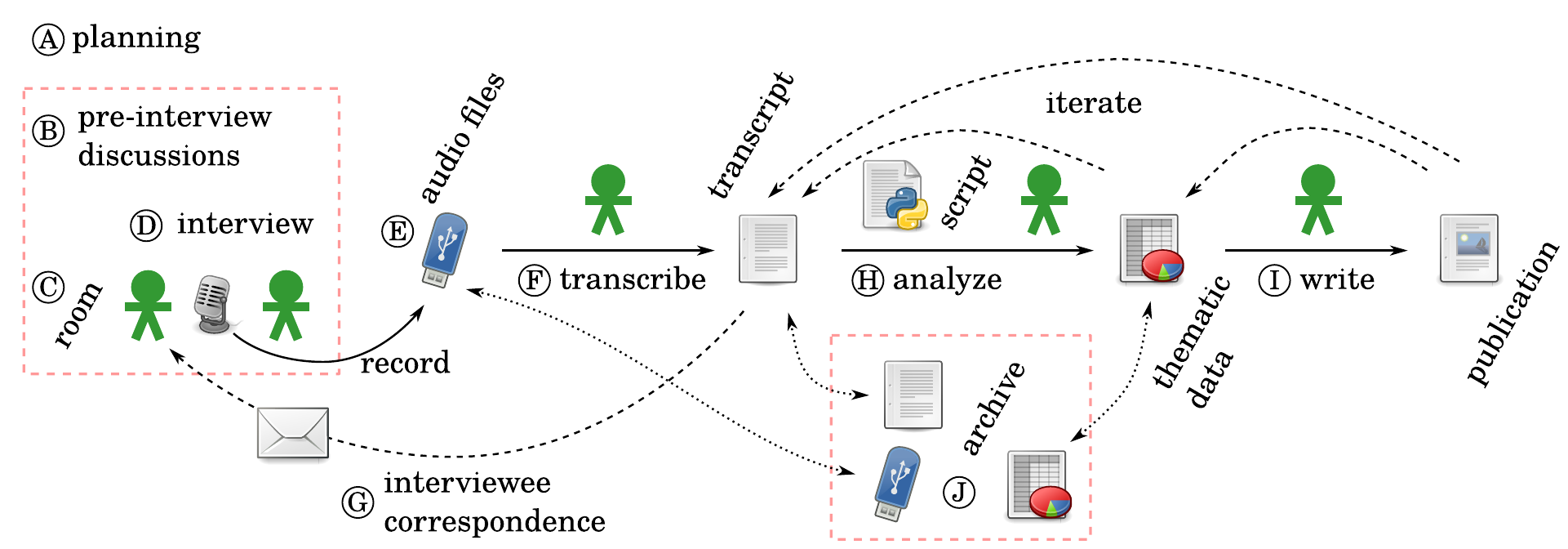}
  \end{center}
\caption{
  Overview of some of the main interview activities (horizontal text)
  and artifacts (sloped text).
  Letters in circles refer to subsections in
  Section~\ref{interview-life-cycle}.
\label{fig-interview-flow}
}
\end{figure*}

\subsection{Legislation and/or Ethics?}
The scope of legislation and ethics is
enormous, and in this paper we will barely scrape the surface of this topic.

In a recent paper by Vardi, he argues that the number of
deaths from automobile crashes has not been decreased with ethics
training for drivers, but with laws and regulations \cite{vardi2019}.
This could imply that we need more laws to govern how researchers
handle interview artifacts.
One such law is the EU General Data Protection Regulation (GDPR)~\cite{gdpr}.
It has almost certainly had an impact on how
interview studies are conducted in the EU since its adoption in 2018.
According to
Schaar, data anonymization, or pseudonymization, are two ways to
comply with GDPR \cite{schaar2016}.

There may be a conflict between ethical guidelines and law.
The declaration of Helsinki argues that no
legal requirement should reduce any of the protections for participants.
However, the declaration does not take precedence over national
law, as was tested in the Gillberg v.\ Sweden trial in the European
Court on Human Rights (after research data had been destroyed
in order to protect interviewees) \cite{gillbergvssweden}.

\subsection{Institutional Review Board}

Many of the guidelines discussed in the above sections point
to the merit of an institutional review board
(IRB\footnote{Alternative terms: research ethics board or independent ethics committee.}).
The Menlo report argues that many researchers in the ICT field
do not know when they are involved in `human subjects research,'
or do not know that this may require involvement of an IRB,
so they do not interact with an IRB at all \cite{menlo2012}.
An ethics board is a great
complement that may aid a researcher into conducting more ethical research.
However, a board is no excuse for a researcher to not,
on his or her own initiative, strive for ethical research.
A board cannot be seen as a catch-all solution for ethical interview studies.
Buchanan et al.\ \cite{buchanan2011},
could be read as a starting point on the topic of IRB's
in computer science security research.

\section{Interview Life Cycle}
\label{interview-life-cycle}

An interview study involves many artifacts, activities, and stakeholders:
the study needs to be planned;
the interviewees have to consent;
the audio needs to be recorded during an interview in a room somewhere;
the audio files have to be archived, transcribed and anonymized;
the transcripts have to be analyzed;
and finally a publication is written.
This might seem like a set of linear activities, but this research is iterative,
and can be time consuming -- we recently finalized a study that required more than two years
from planning to publication.
In the coming sections we:
(i) cover
the elements of an interview study
based on experiences and recommendations from, in particular,
Aldridge et~al.\ \cite{aldridge2010},
Becker-Kornstaedt \cite{becker2001},
Carusi and Jirotka \cite{carusi2009},
and our study:
Strandberg et~al.\ \cite{strandberg2019}.
The interview process is summarized in Figure~\ref{fig-interview-flow}.
(ii) We also introduce a running example with a fictional research project
on the topic of software quality for embedded systems. It involves four individuals:
Alice is a doctoral student, and Bob a post-doc.
Their supervisor, Professor Carol, shares her time
between a nearby university and a helicopter manufacturer.
Professor Carol collaborates with the company contact, Manager Dan.
Finally, (iii) the coming sub-sections contain checklists for the main
elements of an interview study.

\subsection{Planning for Ethical Research}

\myexample{
  Alice, Bob, Carol and Dan are planning for a face-to-face semi-structured
  interview study at the local helicopter company.
  Alice created a draft interview instrument with ten questions
  that she emailed to Bob and Carol, who suggested adding three more questions.
  Manager Dan helped them recruit interviewees.
}

\noindent
\emph{Identify stakeholders:}
In order to achieve the key ethical principle of beneficence,
we should identify stakeholders prior to our research.
Without knowing who the stakeholders are we cannot consider the
potential harm or benefits from a research activity
\cite{mustajoki2017}.
Some obvious stakeholders in software engineering research are:
the interviewee,
the company or organization at which the interviewee is employed,
the researchers conducting the interview,
colleagues of the interviewee (e.g., managers),
other researchers involved in the study
(e.g., supervisors, students recruited for transcription),
industrial practitioners that might benefit from the research results,
the research community in the field of research,
companies whose software is used in the analysis,
(e.g., Google, Microsoft, etc.) and IT-administrators.

The ethical principle of beneficence involves
striving for minimized risk of harm to all stakeholders.
Harm may, in addition to physical harm, involve
risks for the social standing and status in
the family, at work or in the wider community;
risks to privacy and emotions;
as well as risks of revealing information related to
illegal, sexual or deviant behaviour~\cite{carusi2009}.
From a company perspective, harm could be to disclose
intellectual property or
shortcomings in the software development and software quality
processes.
For a practitioner, harm could be caused by revealing
poor performance and non-compliance to
processes to managers or other colleagues \cite{becker2001}.
During interviews, there might be close bonds and trust between the researchers
and the interviewee, and thereby the views of the researchers
might be embedded in the data.
This could be a form of researcher bias, but might also
make the researchers themselves vulnerable to harm
\cite{carusi2009}.
From a wider perspective, harm could also be to produce invalid
research results leading to distrust in certain methods or tools.
In a software engineering context, one could imagine distrust of
test-driven development or C++ as a result
of fraudulent or erroneous research.

\emph{Ethical challenges:}
A second critical step in the planning of ethical interviews
is to recognize that there are ethical challenges
\cite{mustajoki2017}.
Becker-Kornstaedt identifies ethical challenges in the domain of
software process modeling \cite{becker2001},
e.g., managers unexpectedly being present during interviews,
processes not being followed,
the dilemma of de-anonymizing participants or having to obscure data,
as well as dealing with information given ``off the record.''
Other ethical challenges
involve
getting informed 
consent from participants in
research projects;
they might feel forced into participating.
A researcher might also be tempted into
using poor scientific methods instead of using well-established ones
(if the method is flawed
``the results will be invalid so the merit of the study is nil'') \cite{vinson2008}.

\emph{Decisions on ethics:}
When stakeholders and ethical challenges are identified,
a researcher should make decisions on ethics,
and strive for following these decisions,
even when under pressure
\cite{mustajoki2017}.
Alice and Bob might decide they want to anonymize the interview data
in order to protect the interviewees.
One type of pressure comes from the data collection itself:
How should they react if an interviewee mentions critical bugs
in the control system of a helicopter currently being sold?
Another type of pressure comes from the research team:
Would it be unethical of Alice, Bob and Carol
to recruit students to transcribe for extra credits?

\emph{Validate instrument:}
In order to strive for a scientific value of the study,
researchers should validate the instrument.
One way is to do pilot interviews,
other options include expert reviews, focus groups,
cognitive interviews, and experiments \cite{linaker2015}.
Data from pilot interviews are not always used in the final data analysis.

\emph{Involve an IRB:}
During the planning of research involving humans, a researcher should
aim at involving an IRB that could comment on, guide, and/or
approve the research project.

\emph{Our experiences:}
In our study we wrote a research plan following the guidelines
by Linåker et al.\ \cite{linaker2015}.
We made strong commitments to protect the interviewees and the companies,
to anonymize the interviews,
and to destoy data after use.
We conducted three pilot interviews,
after which we did only minor changes to the instrument,
and ended up using two of the three pilot interviews
for the data analysis.
However, we did not carefully identify all stakeholders, we made no harm/benefit analysis,
and we did not involve an IRB.

Alice and Bob did not identify stakeholders,
consider ethical challenges in their research,
validate their instrument, and they did not involve an IRB.


\mychecklist{Planning for Ethical Research}

\begin{enumerate}
\item Are stakeholders identified?
\item Are ethical challenges considered?
\item How will the challenges be addressed?
  Do sponsors and supervisors agree?
\item How will the instrument be validated?
\item Has an IRB been consulted?
\end{enumerate}

\subsection{Pre-Interview Discussions}

\myexample{Alice and Bob inform the interviewees about the purpose and
  topics of the interviews.
  One interviewee mentions that he does not really want to participate in the study,
  but he is worried what Carol and Dan might think if he would not participate.
  After a pep-talk, he gives a really valuable interview.
}

\noindent
Before starting the interview, there must be a discussion with
the interviewee on consent and withdrawal.
These two principles
have been echoed in ethical guidelines since the Nuremberg code.
There should also be a discussion on the purpose and topic of the interview,
as well as a harm-benefit analysis.
Kitchenham et al.\ suggests that interviewees should sign a consent
form, and that this might be needed for an ethical approval
(signing may be clicking a button on a web page)
\cite{kitchenham2015}.
Similarly, for research involving interviews that are funded through
EU Horizon 2020,
informed consent and information sheets are required \cite{eudata2018,euguide2019}.
It may be suitable to archive consent forms, as these may
later be requested by interviewees, funding agencies or
authorities auditing the research quality or data protection.

In Appendix E of Runeson et al.\ \cite{runeson2012}, there is an
example of a consent information letter. It informs the interviewee
about who the researchers are and how to contact them, and it
also highlights that participation is voluntary, the interviewee may
refuse to answer questions and withdraw from the study at any time.
They also inform the interviewee that the interview data will be
protected by law (however, this law has since been outdated and replaced).
Furthermore, the
authors claim that the interview data will be kept confidential
and only available to the research team,
``or in case external quality assessment takes place, to assessors
under the same confidentiality conditions.''
Researchers should never promise that nobody outside
of a research group will ever get access to collected data.
However, many researchers promise this
out of ignorance or because of a mix-up of important terms
\cite{stafstrom2017}.

Saunders et al.\ discuss how anonymized interviewees
could be de-anonymized if the participants are active in social media,
or part of court cases \cite{saunders2015a}.
They recommend to discuss this with the interviewees,
inform them on how data will be anonymized,
and that it might be possible for someone who learns about the interviewee
from multiple sources to de-anonymize him or her.
The participation must be on the basis that the interviewee
understands and accepts this risk.

Understanding what an interviewee consents to can be hard.
The informed consent should be comprehensive,
in a plain language, in the preferred language of the interviewee,
and be accompanied with a discussion
between the researchers and the interviewees
to improve the comprehensibility
\cite{badampudi2017}.

Conducting research without consent 
can be motivated under certain circumstances,
and an IRB may allow deception when:
(i) there is no more than minimal risk to participants,
(ii) the research will not adversely affect the rights and welfare of the
participants,
(iii) the study could not practically be carried out without deception, and
(iv) the participants will debriefed after the study%
\footnote{In addition to interview studies where
interviewees are deceived, we would like to mention two additional research scenarios
where informed consent from every participant may be hard or impossible to collect.
The first scenario involves a researcher studying a criminal bot-net
that has taken control of thousands of smart refrigerators. It is probably not reasonable
for the researcher to collect consent from every owner of an infected refrigerator
to study the impact of the bot-net (example adapted from \cite{menlo2012}).
The second scenario could involve sentiment analysis of bug report discussions in open source projects.
This is a common data collection method for software repository mining research,
and data could be seen as ``publicly available.''
However, the individuals contributing to open source projects
have not given consent to be part of a research project.
These scenarios are not easily translated into an interview study,
and we will not further cover them in this paper.}%
\cite{belmont1978,commonrule}.

\emph{Our experiences:}
In our study \cite{strandberg2019}, we
had a discussion with each interviewee around a written instruction.
The topics included were:
purpose, duration, sampling, sponsor, confidentiality, contact details, and
project leaders,
and were inspired by guidelines from Linåker et al.\ \cite{linaker2015}.
We gave a printed copy of the text to each participant.

Alice and Bob violated the ethical principle of consent by coercing one of the interviewees into participating.

\mychecklist{Pre-Interview Discussions}


\begin{enumerate}[resume]
\item How will informed consent be obtained?
\item How will any participant withdrawals be handled?
\item Are the interviewees informed about purpose,
  possible positive outcomes, possible harm,
  expected duration, sampling, sponsor, confidentiality, contact
  details, project leaders, etc.?
\item What promises, with respect to third party access to interview data, will be made?
  Is there a plan for a potential research quality audit?
\end{enumerate}

\subsection{Room}

\myexample{Alice and Bob got help from manager Dan to book the
  best conference room at the company for the interviews.
  This room has a fancy glass door and is next to the
  most popular coffee machine at the company.
}

\noindent
The room in which the interview is conducted could cause harm to the
stakeholders.
For internal anonymity it should not be obvious
to colleagues that an interview has taken place, nor
what was mentioned.
Before leaving the room, researchers should
remove notes on whiteboards and collect any papers left.
Another ethical risk is when a superior
enters the room and wants to listen to the interview
\cite{becker2001}. This should be avoided
to ensure that the interviewee can speak freely, and to
avoid reactive bias.

\emph{Our experiences:}
For one of our interviews, we had not booked the room long enough,
and at this organization the rooms were in short supply,
so we had to finish the last part of the interview in a
lobby.
This could obviously have broken internal anonymity, and
our discussions could have been overheard by colleagues.

Alice and Bob used a room in plain view to others
which might break internal anonymity
because the colleagues of an interviewee
would know that he or she has been interviewed.

\mychecklist{Room}


\begin{enumerate}[resume]
\item How will internal anonymity be addressed?
\item Are managers informed that their participation might have a
  negative impact on the research?
\item Are interview artifacts removed after interviews?
\end{enumerate}

\subsection{Interview}

\myexample{Alice and Bob find it interesting
  that the helicopter company is a very diverse
  work place.
  In order to capture this in the data, they
  extend the instrument to
  include questions on
  ethnicity, political and religious affiliation, sexual orientation and
  membership in trade unions.
}

\noindent
Before conducting the interviews,
in order to adhere to the ethical principle or researcher skill,
the researchers should have knowledge of, and skills in, research methods in general,
and interview methods in particular \cite{acm2018, nuremberg1947}.
The skills of the researcher will have an impact on the quality
of the interviews, and researchers should also be qualified in
the topic of the interview \cite{eldh2013, hove-anda-2005}.

The two ethical principles of consent and scientific value
should be considered during the interview.
An interviewee who has given consent for research of one purpose
has not given consent for another.
Researchers should take great care to only
collect data that matches the purpose of the research
\cite{eudata2018,euguide2019}.
Data minimization involves limiting the amount of collected
data, reducing the purpose for which it is used, and the period the data is kept.
Data minimization is a central topic in GDPR
but also relevant outside of the EU \cite{eudata2018}.

\emph{Our experiences:}
In our study we did not record the pre-interview discussions.
This way interviewees are not recorded without knowing how the
audio is going to be used,
less audio will make transcription faster, and in case
of a data leak there would be less information lost.

During our interviews we had two researchers present
for most of them;
one took the role of driving the interview and the other kept track of
time and made sure that all questions were asked. The interviews
were semi-structured and we gave the interviewees room to explain
details or complain on problems.
Hove and Anda reported that being two researchers instead of one
seems preferable -- more follow-up questions are asked and more data is recorded \cite{hove-anda-2005}.
%
%

Alice and Bob ask their interviewees about sensitive topics such
as sexual orientation and membership in trade unions.
This is data out of the scope for their research
and they are violating the principle of data minimization.

\mychecklist{Interviews}


\label{todo-interview}

\begin{enumerate}[resume]
\item Do the researchers have adequate skills? 
\item How will data minimization be addressed?
\end{enumerate}

\subsection{Audio Files}

\myexample{During one of the interviews, a participant
  requested a copy of the audio file to be sent to him.
  Alice had recorded the audio on her smartphone and sent it as an
  attachment from her personal email account that was already configured
  in the phone.
}

\noindent
Aldridge et al.\ report on experiences of having
a laptop with sensitive data stolen from the home of a field worker.
They suggest to use a central server with encrypted
connections in order to decrease the number of copies of the
data \cite{aldridge2010}.
They recommend to use passwords for log-in on computers,
and to not allow a computer to remember passwords.
They also recommend encrypting files
so that no one could listen to the audio without decrypting it first.
Furthermore they recommend to make backups, manage the storage and deletion
of data, as well as deleting data permanently when done with it.
Similarly, the Menlo report suggests to destroy risky data
when the research activities are completed (or terminated),
since the data is at risk for as long as it exists \cite{menlo2012}.
This is of particular importance for the audio files
since they contain data that is not yet anonymized.

\emph{Our experiences:}
During our interviews we used an off-line digital voice recorder that
recorded audio as MP3-files. These were stored on a limited number of
computers as well as on a USB-stick in a locked area.
Our motivation for using an off-line digital voice recorder
instead of a smartphone was out of fear that the smartphone producer
would use the data in ways we would not be able to control,
and make backups of the audio files in ways which would render
data ``undeletable.''

We promised the interviewees that we would delete all the
audio files and the links to participants upon publication of the
first paper from the study.
We took a different approach than Aldridge et al.\ and
made sure to \emph{not} store audio files on central servers
or in cloud storage, out of fear that audio would be rendered undeletable.
We also renamed the files to avoid time stamps in the file names, and
we tracked the participant to audio file link on paper only.

Alice and Bob recorded the interview on a smartphone
and emailed the file from a personal email account.
There is thus an obvious risk that the audio will be rendered undeletable
due to backups by the phone manufacturer, or the email provider.
The spread of data that is not yet anonymized
may also cause greater harm than the spread of anonymized data.

\mychecklist{Audio Files}


\begin{enumerate}[resume]
\item What is the data storage plan? \label{data-storage-plan}
\item Has the number of people with access to data been limited?
\end{enumerate}

\subsection{Transcription}

\myexample{Alice and Bob divided transcription work among themselves.
  They transcribed half of the interviews each, and did a round
  of quality control on the other half.
  Bob found it very time consuming to transcribe so he recruited
  two students to do the transcription of the last couple of
  interviews.
}

\noindent \emph{Who Transcribes:}
The transcription process might seem time consuming.
In our study, about a work day was required to transcribe
one hour of interview. In addition, we did a second round of listening
to the audio for quality control of the transcription and the anonymization.
In rough terms we needed 10-15 hours of work to fully
transcribe one hour of audio.
Hove and Anda reported spending about a work day per hour of audio \cite{hove-anda-2005}.
In comparison to the duration of the entire study (more than two years)
time for transcription is not a limiting factor.
Furthermore, the transcription process brings new insights,
and makes the researcher familiar with the data.
We therefore recommend for transcription to be done
by the researchers themselves,
(as do many others, e.g., Runeson et al. \cite{runeson2012}).
Despite this, we recruited students to do
transcription for us.
This forced us to learn about non-disclosure agreements (NDA).
With the help of the University, we had to create an NDA,
we had to get the students to sign the NDA,
and finally we had to archive the signed NDA.
By letting students do transcription, we also
increased the risk of spreading the raw data.
The transcripts we got from the students were also of a lower quality
than the ones we transcribed ourselves. One of the reasons for this was
lack of familiarity with the domain-specific jargon
-- Lethbridge et al.\ had similar experiences \cite{lethbridge2008}.
This led us to do fill-in transcription, corrections and additional
anonymization.

\emph{What to Anonymize:}
Surmiak
interviewed 42
researchers in the field of sociology and anthropology
who, in turn, do research with vulnerable participants
(sexual minorities, homeless, war veterans, etc.) in Poland \cite{surmiak2018}.
She wanted to know how researchers manage confidentiality.
She
sent the transcripts to the interviewed researchers
for corrections, clarifications and further anonymization.
Surmiak
found that other researchers were very aware of the risk
of being de-anonymized from the transcripts and many wished to
review their transcripts -- 
the researchers wanted to be treated
in another way than they, in turn, treat their participants.
This might come from the awareness the researcher has,
and that they might be less trusting towards other researchers.

Vinson and Singer 
mention three principles of confidentiality:
(i)
data privacy: limit access to the data,
(ii)
data anonymity: examination of data should not lead to de-anonymization,
and
(iii)
and anonymity of participation
(or internal anonymity): participation is not revealed to colleagues
\cite{vinson2008}.
Runeson et al., just like Kitchenham et al.,
suggest that companies and individuals could be de-anonymized
with too many details, or a too small sample,
\cite{runeson2012, kitchenham2015}.
Becker-Kornstaedt suggests to
interview more than one person per role, project or department
as a possible mitigation
\cite{becker2001}.
During the transcription we recommend to do anonymization
while listening, but before writing;
this way sensitive information will never be saved to disk in plain text.
Saunders et al.\ suggest anonymizing
people's names, places, religious or cultural background,
occupation, family relationships, and other potentially identifying
information \cite{saunders2015b, saunders2015a}.
Surmiak also mentions:
occupation, place of work, nationality, religion, hobbies,
military rank,
gender, zodiac sign, dietary restrictions, and periods of illness
\cite{surmiak2018}.

In our study \cite{strandberg2019} we 
anonymized about 90 extracts per interview.
The three most anonymized categories were:
313 extracts of tools or tool jargon
(e.g., programming language, and version control system),
178 names or details of products and organizations, and
160 extracts related to the domain.
Other categories we anonymized were
company specific jargon,
technical details,
names of places and people,
numbers or points in time,
and
off-topic discussions.

\emph{How to Anonymize:}
There seem to be two major approaches to anonymization.
One is to assign pseudonyms. In some cases it may be motivated to assign
more than one pseudonym to one interviewee; Saunders et al.\
did this when one extract of a transcript will not
identify an interviewee, but when the combination of two might.
They also mention approaches where multiple interviewees are assigned
to one pseudonym in order to create a more representational story.
Several papers highlight that keeping track of pseudonyms can
be hard when the number grows --
in software engineering
this could for example
happen if researchers are discussing a number of subsystems
and their interfaces when every subsystem has a pseudonym and
a number of people working with them.

We anonymized interviews by replacing some words with more general terms
within pointy brackets.
For example, C++ would be <programming language> and
helicopter would be <vehicle>, etc.
We listened to the audio in a media player running at low speed and
wrote the transcript in plain text files with
speech from researchers, and interviewees on separate lines
with an initial ``Q'' for questions or comments from the researchers,
and ``A'' for answers by the interviewee
(Saunders et al.\ instead transcribe with ``Interviewer''
and a pseudonymized name of the interviewee).
Pauses were indicated with blank lines followed by a
time\-stamp showing the number of minutes and seconds into the recording.
This way, a researcher could easily go back to the original
recording if a transcript appeared incorrect.
Example transcript:

\begin{quote}
[28:54] \\
Q: The next part is on testing and test results. We've covered some of
this perhaps.  Err\ldots\ But could you give an example of a typical
test case? \\
A: Actually, we should have a look into, into <requirements management
tool> to see what it looks like. But I mean, for example a <vehicular
mechanism> sequence.
\end{quote}

\noindent
The same answer transcribed in the style of Saunders et al.:
\begin{quote}
  Actually, we should have a look into Req\-Test\-Tracker
  to see what it looks like.
  But I mean, for example
  the safe full stop for maintenance sequence.
\end{quote}

\noindent
Just like we did, Alice and Bob recruited students to do the transcription.
We recommend to do transcription within the research team.

\mychecklist{Transcription}

\begin{enumerate}[resume]
\item Who will transcribe the audio?
\item How will meta information (such as separation of speakers, timestamps, etc.) be added to the transcripts?
\item How will consistent transcription over interviews, and over researchers, be achieved?
\item What will be anonymized?\footnote{
  Candidates for anonymization are:
    names of people, places, companies, organizations, tools, and products;
    domain-specific details such as
    programming languages,
    domain-specific terminology,
    company specific jargon and technical details that are not of relevance
    to the topic of the interview;
    numbers and points in time such as birthdays,
    graduation years, number of years in a work place,
    number of colleagues, number of subsystems in a product,
    or
    number of lines of code in a product;
    personal details such as religion, cultural background,
    military rank, hobbies, nationality,
    occupation, family relationships, gender,
    zodiac sign, dietary restrictions, periods of illness, etc.;
    and also
    off-topic details, such as the pre-interview discussion,
    or a rant from the interviewee.
  }
\end{enumerate}

\subsection{Interviewee Correspondence}

\myexample{With the exception of the interviewee that requested
  an audio file, Alice and Bob never contacted the
  interviewees again. ``If they were interested in the results
  of the study, then they could read the paper once it's out,''
  they argued.
}

\noindent
By corresponding with the interviewees they can be given the opportunity to review, 
correct, clarify or expand on the interview.
Surmiak gave her interviewees the chance to not only review,
but also to rewrite transcripts
\cite{surmiak2018}.
A review step is recommended by e.g.\ Runeson et al.\ \cite{runeson2012}.

\emph{Our experiences:}
We gave the interviewees the opportunity to review the transcripts
and expected that some of them might have
wanted to comment on or clarify something.
At the end of the interview, we asked the
interviewees if they wanted a copy of the transcript, and if so, in
which format they wanted it.  For maximum anonymity, we had expected
them to want it on physical paper, possibly sent to their home
address. However, those that wanted to review the transcript
wanted it sent by email, and all except one wanted it to their work
address.
This, obviously, makes it possible for an IT-department at their
companies (and the organization from which it was sent)
to read the transcripts and to link them to individuals interviewed.
It might also render the transcript undeletable.
It is not clear that the interviewees understood this risk,
however, we complied in their request to send it by email to them.

Alice and Bob did not give their interviewees the possibility to review the transcripts.

\mychecklist{Interviewee Correspondence}


\begin{enumerate}[resume]
\item Will interviewees review transcripts? \label{i-review-t}
\item If yes to \ref{i-review-t}, how is correspondence to be conducted?
\item If yes to \ref{i-review-t}, will they be given the possibility to delete, correct, clarify and/or expand on the transcripts?
\end{enumerate}

\subsection{Data Analysis and Thematic Data}

\myexample{Some time after the interview,
  one of the interviewees requested to withdraw from the study.
  Alice and Bob agreed, and deleted the corresponding audio file
  and transcript, but the thematic data was kept
  since it was hard coded in the scripts,
  already in the spread sheets,
  and the paper was to be submitted the same week.
}

\noindent
For data analysis in studies involving qualitative data,
coding is common. It is suggested for
thematic analysis \cite{braun2006},
content analysis \cite{graneheimlundman},
grounded theory \cite{glaserstrauss, stol2016}
etc. In coding, some parts of a transcript are ``tagged''
with codes and themes in various hierarchies,
making it possible to understand and
investigate the data in different ways. Commercial tools
for coding are available.
%
If an on-line spread sheet like Google Docs is
used in research, the researchers must be aware
of the old saying with roots from the 1970's:
``if you are not paying for it, you're not the customer,
you're the product being sold''
\cite{payproduct}.
Indeed, according to the Google Safety Center,
Google will collect
``Docs, Sheets, and Slides you create on Drive''
and use it ``to make Google services more useful for you''
\cite{googlesteal}.
What this means in terms of interview anonymity, and
data longevity is unclear, but a researcher should take great caution
and not store sensitive (i.e., not yet anonymized) information in
these types of tools.

Aldridge et al.\ suggests to only use `on screen' working methods,
and when not doing so, the paper copies should immediately be shredded
after use \cite{aldridge2010}.

\emph{Our experiences:}
Prior to data analysis, we double checked the anonymization of the transcripts.
We analyzed the data with thematic analysis in on-line spread sheets,
and also shared anonymized data with a cloud storage service.
The removal of an interview would have been as easy as removing lines in a spread sheet,
and would have had no impact on the scripts analyzing the data.

Alice and Bob only partially deleted data once an interviewee wanted to withdraw.
Again, they violate the ethical principle of consent, and possibly also
the principles of confidentiality and respect for law.

\mychecklist{Data Analysis and Thematic Data}


\begin{enumerate}[resume]
\item Will data analysis (and the potential use of third party tools) be done on anonymized data only?
\item Has the end user license agreements for tools been read?
\item Is there an inventory of the data (with locations of audio files, transcripts,
and processed data)?
\end{enumerate}

\subsection{Writing and the Publication}

\myexample{Alice and Bob want to protect the interviewees
  and hide the name of the company in the paper, and describe it as
  ``a Nordic manufacturer of manned helicopters with about 1500
  employees.''\footnote{%
    This is a fictional example, as far as we know
    there are no manufacturers of manned helicopters
    with 1500 employees in any of the Nordic countries.
    }
}

\noindent
One of the drawbacks of using anonymized data
is the lack of context. Results from
``an embedded systems company'' may convey a
different meaning than results from ``a Nordic helicopter manufacturer
recovering from a series of bribery scandals with a next generation
helicopter that will make or break the company.''
Reporting on context is important in empirical software
engineering \cite{petersen2009context}.
However, if the paper contains too many details on context,
then the companies involved could be identified.
Therefore, when reporting on context
one should report only on organizations and interviewees in an
aggregated form.

Before publishing any paper,
third party reviewers
might need to revisit or make an audit of the data process.
Researchers should therefore have the data in order, and be able to
explain the flow from plan to paper.

In order to foster trust and communication between industry and academia,
and to honor the ethical principle of justice,
it is important to give feedback to the participants and
the organizations that were part of the study,
as well as other industrial practitioners and society as a whole.
%
For the research
to better reach these groups, it may be motivated for the researchers
to publish papers with an open access license,
to return to the companies with presentations or reports in
formats other than typical academic papers, or to make
video recordings of presentations.
If knowledge is not given, then other gains could be considered \cite{eldh2013}.

Finally, there are a number of recommendations on what to include
in an academic paper:
(i) 
how the interviews
were conducted and, if possible, include the questions asked \cite{singer2008},
(ii) 
ethical aspects,
such as how consent was received \cite{badampudi2017},
(iii) 
context
\cite{petersen2009context},
(iv) Runeson et al.\ \cite{runeson2012} propose
additional topics such as validity,
and, as is common in most academic papers,
(v) a section on method
in order to let a reader know that the research is sound --
without a valid method the results could be meaningless.

\emph{Our experiences:}
We reported on context in an aggregated form only: e.g.\ details on
company size was kept separate from details on domain, etc.
%
Alice and Bob report on context in a way that may uniquely identify the company.

\mychecklist{Writing and the Publication}


\begin{enumerate}[resume]
\item How will details on the organizations, and other context data, be reported?
\item Will reports in different forms, for different audiences, be prepared?
\item How will feedback for the participating interviewees and organizations be made?
\end{enumerate}

\subsection{Archive}

\myexample{Professor Carol got a new position at a more prestigious
  university and left her previous positions. Her new research group
  will focus on agile practices for embedded systems.
  In order to kick-start this research she brought the interview transcripts
  from the helicopter company to the new group
  where this data will be combined with findings from a literature
  study to provide new insights.
}

\noindent
Research data is not the private property of the researchers,
and should not be treated as such.
For both ethical and legal principles,
a researcher should consider both archiving data \cite{stafstrom2017},
and destroying data from the archive \cite{menlo2012}.
Some of the reasons for archiving data are
(i) to support investigations of scientific misconduct,
(which was an important topic in the Gillberg v.\ Sweden trial \cite{gillbergvssweden}),
(ii) data re-use by the researchers themselves or others,
and (iii) if the data is of general importance to society at large,
it could have value, in itself, for coming generations \cite{stafstrom2017}.
Practical advice on what and how to archive are missing in
standard popular software engineering research guidelines such as
Kitchenham et al. \cite{kitchenham2015},
Runeson et al.\ \cite{runeson2012},
and Shull et al.\ \cite{shull2008}.
Researchers, even in software engineering, should
also be aware of the existence of laws regulating
archiving.

Carusi and Jirotka investigated archiving of
qualitative data.
They argue that
de-anonymization may be trivial, in particular when dealing with
body language and facial expression data
\cite{carusi2009}.
When dealing with data of new types, there is
often a lack of ethical guidelines, and there might be a conflict
between requirements from funding agencies, the academia and laws.
Informed consent may also be impossible when
a participant does not understand the media type.
Finally, withdrawal might be impossible if data is publicly archived.
%
However, allowing a
participant to withdraw is fundamental to ethical research,
so researchers planning on sharing qualitative data should both
strongly anonymize it, and also ensure that the interviewee fully understands
the limitations with respect to withdrawal from the study.

\emph{Our experiences:}
In our study we stored the data in an internal archive during the study.
Upon first publication from the interviews, we destroyed the
audio files as well as links between individuals and transcripts.
Within ten years from the time of the interviews, we will destroy
the remaining data.

Carol took interview data from one research group to another.
It is unlikely that the interviewees gave their consent for this,
and it might also violate the ethical principle of respecting laws.

\mychecklist{Archive}


\begin{enumerate}[resume]
\item If any data is to be publicly archived,
  how will the implications with respect to de-anonymization and withdrawal from the study be explained to the interviewees?
\item What is the data deletion plan? When, how and by whom will the data be deleted? Is it coordinated with the data storage plan (item \ref{data-storage-plan})?
\end{enumerate}

\section{Summary and Discussion}
\label{summary}
Research ethics is a vast field that is hard to get an overview of.
It is difficult to
consider every aspect of stakeholders, activities and
artifacts from an ethical perspective.
In this paper we have reviewed existing guidelines for ethical
research, guidelines for software engineering, and revisited an interview
study recently finalized.
There is a gap in previous work on how to apply ethical principles
for interviews in software engineering research.
We have addressed this gap by considering ethical aspects of each step in an interview study,
and provide check\-lists for these steps.
These check\-lists
give researchers a stable platform for a
more ethical research project.

The check\-lists are based on previous work and our own experiences.
Of particular importance are the previous publications by
Becker-Kornstaedt, who composed a list of ethical challenges
in descriptive software process modeling \cite{becker2001};
Aldridge et al., who listed ways in which an interview is copied,
how it proliferates and suggested a number of guidelines to avoid
spread of sensitive data \cite{aldridge2010};
Saunders et al., who proposed guidelines for anonymization
and experiences from participants active in social media,
or in court cases \cite{saunders2015b, saunders2015a};
as well as
Surmiak, who interviewed interviewers about
confidentiality involving vulnerable interviewees
\cite{surmiak2018}.

All research has limitations. In this study we have investigated a
large number of guidelines with an origin in the field of medicine.
There are, of course, an even larger number of guidelines that we have
not investigated -- both in medical fields such as psychology, and in
fields related to software engineering. It is therefore likely that
the check\-lists proposed in this paper are incomplete. However, we would
like to see this as a starting point for researchers unsure of
the ethics in their interview research,
and we encourage other studies to build upon, revise or reject
our recommendations.

We would welcome future work
on how to make an interview study that is compliant with
the increasing complexity of
laws and regulations, such as how to comply with GDPR, the Helsinki
declaration, national or international laws, and at the same time
fulfill requirements by sponsors.

In software engineering research, the use of an IRB seems
immature despite recommendations in guidelines such as the Menlo
report. Future work could investigate at which level there are legal
requirements on empirical software engineering to use or start using
an IRB before doing research involving humans,
as well as guidelines on how to get started with an IRB
at a university where there is no such board.

A third possible field of future work involves the knowledge and
competence of the researchers themselves.
Researchers in the field of
software engineering should follow best practices
with respect to research methods.
Guidelines, e.g.\ the books by
Kitchenham et al. \cite{kitchenham2015},
Runeson et al.\ \cite{runeson2012},
and Shull et al.\ \cite{shull2008}
instruct a researcher on how to do research.
We would welcome research providing instructions;
e.g.\ a check\-list for researchers, supervisors, or reviewers;
on when \emph{not to} conduct research,
such that we may avoid doing research blindly.

\section{Conclusion}
\label{conclusion}

Despite laws and regulations, research ethics is a hard and
unnatural topic for many empirical
software engineering researchers.
In this paper we have learned from our own experiences
and listened to the authority of existing guidelines, in order
to distill a comprehensive guide for interview studies.
In particular, we suggest how to hands-on anonymize
interview data in the transcription process.
This gives researchers a platform
for a more ethical research project.

\section{Acknowledgments}
  This work was sponsored by Westermo Network Technologies AB,
  and the Knowledge Foundation through the grants 20150277 (ITS ESS-H)
  and 20160139 (TestMine).

  The author would like to thank
  Adrianna Surmiak,
  Aida Causevic,
  Tom Ostrand,
  Wasif Afzal,
  Daniel Sundmark,
  and
  Elaine Weyuker
  for valuable discussions during the writing of this paper.


\begin{thebibliography}{99}
\begin{small}

\bibitem{aldridge2010}
J Aldridge, J Medina, and R Ralphs.
``The problem of proliferation: guidelines for improving the security of qualitative data in a digital age.''
\emph{Research Ethics}, 6(1), 3-9, 2010.

\bibitem{badampudi2017}
D Badampudi.
``Reporting ethics considerations in software engineering publications.''
In \emph{International Symposium on Empirical Software Engineering and Measurement (ESEM)}, 2017.

\bibitem{becker2001}
U Becker-Kornstaedt.
``Descriptive Software Process Modeling -- How to Deal with Sensitive Process Information.''
\emph{Empirical Software Engineering}, 6(4), 353-367, 2001.

\bibitem{braun2006}
V Braun, and V Clarke.
``Using thematic analysis in psychology.''
\emph{Qualitative research in psychology}, 3(2), 77-101, 2006.

\bibitem{buchanan2011}
E Buchanan, J Aycock, S Dexter, D Dittrich, and E Hvizdak.
``Computer science security research and human subjects: Emerging considerations for research ethics boards.''
\emph{Journal of Empirical Research on Human Research Ethics}, 6(2), 71-83, 2011.

\bibitem{caplan2011}
A Caplan.
``The use of prisoners as sources of organs -- an ethically dubious practice.''
\emph{The American Journal of Bioethics}, 11(10), 1-5, 2011.

\bibitem{carusi2009}
A Carusi, and  M Jirotka.
``From data archive to ethical labyrinth.''
\emph{Qualitative Research}, 9(3), 285-298, 2009.

\bibitem{commonrule}
Department of Homeland Security et al.\
``Federal policy for the protection of human subjects.''
\emph{Federal Register}, 82(12), 7149–7274, 2017.

\bibitem{menlo2012}
D Dittrich, and E Kenneally,
``The Menlo Report: Ethical principles guiding information and communication technology research.''
US Department of Homeland Security,
2012.

\bibitem{menlo2013}
D Dittrich, E Kenneally, and M Bailey.
``Applying ethical principles to information and communication technology research: A companion to the Menlo Report.''
US Department of Homeland Security,
2013.

\bibitem{eldh2013}
S Eldh.
``Some researcher considerations when conducting empirical studies in industry.''
In \emph{International Workshop on Conducting Empirical Studies in Industry (CESI)}, 2013.

\bibitem{eudata2018}
European Commission.
``Horizon 2020 Online Manual: Ethics and data protection.''
DG for Research and Innovation, 2018.
\href{http://ec.europa.eu/research/participants/data/ref/h2020/grants_manual/hi/ethics/h2020_hi_ethics-data-protection_en.pdf}{Online},
accessed 2019-03-27.

\bibitem{euguide2019}
European Commission.
``Horizon 2020 Online Manual:  Guidance How to complete your ethics self-assessment.''
DG for Research and Innovation. V.\ 6.1, 2019.
\href{http://ec.europa.eu/research/participants/data/ref/h2020/grants_manual/hi/ethics/h2020_hi_ethics-self-assess_en.pdf}{Online},
accessed 2019-03-27.

\bibitem{gillbergvssweden}
European Court on Human Rights,
Grand Chamber judgment Gillberg v. Sweden,
application no.\ 41723/06,
Press Release, 2012.
\href{https://hudoc.echr.coe.int/app/conversion/pdf?library=ECHR&id=003-3901658-4502025&filename=Grand\%20Chamber\%20judgment\%20Gillberg\%20v\%20\%20Sweden\%2003.04.12.pdf}{Online},
accessed 2019-03-27.

\bibitem{gdpr}
European Parliament and Council of the European Union,
``Regulation on the protection of natural persons with regard to the processing of personal data and on the free movement of such data, and repealing Directive 95/46/EC (Data Protection Directive).''
\emph{Official Journal of the European Union}, L 119, p. 1–88, 2016.

\bibitem{helsinki2014}
General Assembly of the World Medical Association.
``Declaration of Helsinki – Ethical Principles for Medical Research Involving Human Subjects.''
\emph{64th WMA General Assembly, Fortaleza, Brazil}, 2013.
\href{https://www.wma.net/policies-post/wma-declaration-of-helsinki-ethical-principles-for-medical-research-involving-human-subjects/}{Online},
accessed 2018-12-13.

\bibitem{taipei2016}
General Assembly of the World Medical Association.
``Declaration of Taipei on Ethical Considerations regarding Health Databases and Biobanks.''
\emph{67th WMA General Assembly, Taipei, Taiwan}, 2016.
\href{https://www.wma.net/policies-post/wma-declaration-of-taipei-on-ethical-considerations-regarding-health-databases-and-biobanks/}{Online},
accessed 2018-12-13.

\bibitem{glaserstrauss}
B G Glaser, and A L Strauss.
\emph{The Discovery of Grounded Theory: Strategies for Qualitative Research.}
Chicago: Aldine,
1967.

\bibitem{googlesteal}
Google Safety Center.
``Making it easy to understand what data we collect and why.''
\href{https://safety.google/privacy/data/}{Online},
accessed 2019-02-13.

\bibitem{acm2018}
D Gotterbarn (chair), and the ACM Code 2018 Task Force.
``ACM Code of Ethics and Professional Conduct.''
ACM, 2018.

\bibitem{graneheimlundman}
U H Graneheim, and B Lundman.
``Qualitative content analysis in nursing research: concepts,
procedures and measures to achieve trustworthiness.''
\emph{Nurse education today,} 24(2), 105-112,
2004.

\bibitem{hippo1923}
Hippocrates of Cos.
\emph{The Oath},
Loeb Classical Library 147,
1923.

\bibitem{hove-anda-2005}
S E Hove, and B Anda.
``Experiences from Conducting Semi-Structured Interviews in Empirical Software Engineering Research.''
In \emph{IEEE International Software Metrics Symposium (METRICS)},
2005.

\bibitem{kitchenham2015}
B A Kitchenham, D Budgen, and P Brereton.
\emph{Evidence-based software engineering and systematic reviews (Vol. 4)}.
CRC press, 2015.

\bibitem{lethbridge2008}
T C Lethbridge, S Lyon, and P Perry.
``The Management of University -- Industry Collaborations Involving Empirical Studies of Software Engineering.''
In \emph{Guide to Advanced Empirical Software Engineering (pp. 257-281)}. Springer, London, 2008.


\bibitem{linaker2015}
J Linåker, S M Sulaman, R Maiani de Mello, and M Höst.
``Guidelines for conducting surveys in software engineering.''
Department of Computer Science, Lund University, Sweden, 2015.


\bibitem{belmont1978}
National Commission for the Protection of Human Subjects of Biomedicaland Behavioral Research, Bethesda, Md.
\emph{The Belmont report: Ethical principles and guidelines for the protection of human subjects of research}.
Superintendent of Documents, 1978.

\bibitem{nespor2000}
J Nespor.
``Anonymity and place in qualitative inquiry.''
\emph{Qualitative inquiry}, 6(4), 546-569, 2000.

\bibitem{mustajoki2017}
H Mustajoki and A Mustajoki.
\emph{A New Approach to Research Ethics: Using Guided Dialogue to Strengthen Research Communities.}
Routledge, 2017.

\bibitem{petersen2009context}
K Petersen, and C Wohlin.
``Context in industrial software engineering research''.
In \emph{International Symposium on Empirical Software Engineering and Measurement.} IEEE Computer Society, 2009.


\bibitem{payproduct}
Quote Investigator.
``You're Not the Customer; You're the Product''
\href{https://quoteinvestigator.com/2017/07/16/product/}{Online},
accessed 2019-02-13.


\bibitem{rogers2019}
W Rogers, M P Robertson, A Ballantyne B Blakely, R Catsanos, R
Clay-Williams, and M F Singh.
``Compliance with ethical standards in the reporting of donor sources
and ethics review in peer-reviewed publications involving organ
transplantation in China: a scoping review.''
\emph{BMJ Open}, 2019.

\bibitem{rosenblum2011}
N Rosenblum, X Zhu, and B P Miller.
``Who wrote this code? Identifying the authors of program binaries.''
In \emph{European Symposium on Research in Computer Security} (pp. 172-189).
Springer, Berlin, Heidelberg, 2011.

\bibitem{runeson2012}
P Runeson, M Höst, A Rainer, and B Regnell.
\emph{Case study research in software engineering: Guidelines and examples}.
John Wiley \& Sons, 2012.

\bibitem{saunders2015b}
B Saunders, J Kitzinger, and C Kitzinger.
``Anonymising interview data: Challenges and compromise in practice.''
\emph{Qualitative Research}, 15(5), 616-632, 2015.

\bibitem{saunders2015a}
B Saunders, J Kitzinger, and C Kitzinger.
``Participant anonymity in the internet age: from theory to practice.''
\emph{Qualitative research in psychology}, 12(2), 125-137, 2015.

\bibitem{schaar2016}
K Schaar.
``What is important for Data Protection in science in the future? General and specific changes in data protection for scientific use resulting from the EU General Data Protection Regulation.''
\emph{RatSWD Working Paper, No. 258}, Berlin, 2016.

\bibitem{shull2008}
F Shull, J Singer, and D I K Sjøberg (Eds.).
\emph{Guide to advanced empirical software engineering.}
Springer Science \& Business Media, 2008.

\bibitem{singer2008}
J Singer, S E Sim, and T C Lethbridge.
``Software engineering data collection for field studies.''
In \emph{Guide to Advanced Empirical Software Engineering} (pp. 9-34).
Springer, London, 2008.


\bibitem{smith2005}
C M Smith.
``Origin and uses of primum non nocere -- above all, do no harm!''
\emph{The Journal of Clinical Pharmacology}, 45(4), 371-377,
2005.

\bibitem{stafstrom2017}
S Stafström.
\emph{God forskningssed [Good research practice]}.
Stockholm, Vetenskapsrådet, 2017.
\href{https://www.vr.se/analys-och-uppdrag/vi-analyserar-och-utvarderar/alla-publikationer/publikationer/2017-08-29-god-forskningssed.html}{Online},
accessed 2018-12-13.

\bibitem{stol2016}
K J Stol, P Ralph, and B Fitzgerald.
``Grounded theory in software engineering research: a critical review and guidelines.''
In \emph{IEEE International Conference on Software Engineering (ICSE)},
2016.

\bibitem{strandberg2019}
P E Strandberg, E P Enoiu, W Afzal, D Sundmark, and R Feldt.
``Information Flow in Software Testing -- An Interview Study with Embedded Software Engineering Practitioners.''
In \emph{IEEE Access, vol.~7}, pp. 46434-46453, 2019.

\bibitem{surmiak2018}
D A Surmiak.
``Confidentiality in qualitative research involving vulnerable participants: Researchers' perspectives.''
In \emph{Forum Qualitative Sozialforschung/Forum: Qualitative Social Research.}
DEU, 2018.

\bibitem{sweeney2000}
L Sweeney.
``Simple demographics often identify people uniquely.''
\emph{Health} (San Francisco), 671, 1-34, 2000.



\bibitem{nuremberg1947}
U.S.\ Government Printing Office.
``Permissible Medical Experiments''
in \emph{Trials of War Criminals before the Nuremberg Military Tribunals under Control Council Law}
No. 10, Vol. 2, pp. 181-182. Washington, D.C. 1949.
\href{https://www.loc.gov/rr/frd/Military_Law/pdf/NT_war-criminals_Vol-II.pdf}{Online},
accessed 2018-12-14.

\bibitem{vardi2019}
M Y Vardi.
``Are we having an ethical crisis in computing?''
\emph{Communications of the ACM}, January 2019, Vol. 62 No. 1, Page 7, 2019.

\bibitem{vinson2008}
N G Vinson, and J Singer.
``A practical guide to ethical research involving humans.''
In \emph{Guide to Advanced Empirical Software Engineering (pp. 229-256).}
Springer, London, 2008.



\end{small}

\end{thebibliography}
\end{document}